\documentclass[11pt]{article}
\usepackage{amsmath,amssymb,color,graphics,epsfig}

\usepackage{amsfonts}

\newcommand{\be}{\begin{equation}}
\newcommand{\ee}{\end{equation}}
\newcommand{\bea}{\setlength\arraycolsep{2pt} \begin{eqnarray}}
\newcommand{\eea}{\end{eqnarray}}
\newcommand{\nn}{\nonumber}

\def\ft#1#2{{\textstyle{\frac{\scriptstyle #1}{\scriptstyle #2} } }}
\def\fft#1#2{{\frac{#1}{#2}}}

\def\0{{\sst{(0)}}}
\def\1{{\sst{(1)}}}
\def\2{{\sst{(2)}}}
\def\3{{\sst{(3)}}}
\def\4{{\sst{(4)}}}
\def\5{{\sst{(5)}}}
\def\6{{\sst{(6)}}}
\def\7{{\sst{(7)}}}
\def\8{{\sst{(8)}}}
\def\sst#1{{\scriptscriptstyle #1}}

\begin{document}

\begin{flushright}
\end{flushright}

\vspace{25pt}
\begin{center}
{\large {\bf Black holes in vector-tensor theories and their thermodynamics}}

\vspace{10pt}
 Zhong-Ying Fan

\vspace{10pt}
{ Center for Astrophysics, School of Physics and Electronic Engineering, \\
 Guangzhou University, Guangzhou 510006, China }\\


\vspace{40pt}

\underline{ABSTRACT}
\end{center}
In this paper, we study Einstein gravity either minimally or non-minimally coupled to a vector field which breaks the gauge symmetry explicitly in general dimensions. We first consider a minimal theory which is simply the Einstein-Proca theory extended with a quartic self-interaction term for the vector field. We obtain its general static maximally symmetric black hole solution and study the thermodynamics using Wald formalism. The aspects of the solution are much like a
Reissner-Nordstr\o m black hole in spite of that a global charge cannot be defined for the vector. For non-minimal theories, we obtain a lot of exact black hole solutions, depending on the parameters of the theories. In particular, many of the solutions are general static and have maximal symmetry. However, there are some subtleties and ambiguities in the derivation of the first laws because the existence of an algebraic degree of freedom of the vector in general invalids the Wald entropy formula. The thermodynamics of these solutions deserves further studies.


\thispagestyle{empty}

\pagebreak

\tableofcontents
\addtocontents{toc}{\protect\setcounter{tocdepth}{2}}




\section{Introduction}
In recent years, the vector-tensor theories (or generalized Einstein-Proca theories) have attracted a lot of attentions in the literatures \cite{Jimenez:2009ai,Jimenez:2008sq,Jimenez:2008au,Tasinato:2014eka,Heisenberg:2014rta,Allys:2015sht,Jimenez:2016isa,DeFelice:2016cri,DeFelice:2016yws,DeFelice:2016uil,Geng:2015kvs,
Chagoya:2016aar,Fan:2016jnz,Minamitsuji:2016ydr,Cisterna:2016nwq,Babichev:2017rti,Heisenberg:2017xda,Heisenberg:2017hwb,Chagoya:2017ojn}. In these theories, there exist some interesting solutions, which are relevant for astrophysics and cosmology, such as the stealth Schwarzschild black hole. The existence of such solutions breaks the uniqueness theorem of spherically symmetric solutions in General Relativity and provides new candidates for astrophysical tests. The cosmological implications of these theories were also studied in a series of literatures \cite{Jimenez:2009ai,Jimenez:2008sq,Jimenez:2008au,Tasinato:2014eka,DeFelice:2016cri,DeFelice:2016yws,DeFelice:2016uil}. On the other hand, there exists a no-go theorem which excludes the existence of Einstein-Proca black holes in asymptotically flat space-times \cite{Bekenstein:1971hc,Bekenstein:1972ky}. However, the theorem is easily evaded. By numerical analysis, it was established in \cite{Herdeiro:2016tmi} that when Einstein gravity minimally coupled to an even number of real Proca fields, there exist asymptotically flat, stationary, axisymmetric black holes with Proca hair. It was analytically shown in \cite{Geng:2015kvs,
Chagoya:2016aar,Fan:2016jnz,Minamitsuji:2016ydr,Cisterna:2016nwq,Babichev:2017rti,Heisenberg:2017xda,Heisenberg:2017hwb,Chagoya:2017ojn} that the no-go theorem can be avoided in the presence of non-minimal couplings between the curvature and the vector fields.

Yet, there are still some holes left in the literatures which motive our current work. The first is in vector-tensor theories the vector field is as physical as the field strength since the gauge symmetry is explicitly breaking owing to either a nonzero bare mass or non-minimal couplings. A direct consequence of this is one can introduce a non-trivial radial component for the vector field $A_r$ when solving black hole solutions in the dual theories \cite{Chagoya:2016aar,Minamitsuji:2016ydr,Babichev:2017rti}. However, the power of this has not been considered very well. In this paper, we will show that in many cases how one can obtain the most general static maximally symmetric solutions with a nonzero $A_r$ in general dimensions. The most simple example we study is a minimal theory which generalizes the free massive Proca theory with a quartic self-interaction term. We also obtain such general static solutions for non-minimal theories with coupling terms of the form $R_{\mu\nu}A^\mu A^\nu$ and $G_{\mu\nu}A^\mu A^\nu$, where $R_{\mu\nu}\,,G_{\mu\nu}$
are the Ricci tensor and Einstein tensor of the metric respectively. In the four dimension, the later case has been well studied in \cite{Babichev:2017rti}.

Our second motivation is
while people have obtained a series of exact black holes with vector hairs, their thermodynamics was even not studied except for a few papers \cite{Geng:2015kvs,Fan:2016jnz}. Here we will adopt the Wald formalism to derive the first law of thermodynamics systematically for all the solutions we obtain. Moreover, we find some subtleties and ambiguities when deriving the first laws for the solutions with a nonzero $A_r$. The underlying reason is $A_r$ is a purely algebraic degree of freedom which does not introduce corresponding vector charges in the solutions. However, to govern the validity of Wald entropy formula, one should impose proper boundary conditions on the horizon for both $A_t$ and $A_r$, which in general results to a degenerate solution characterized by only one parameter, in contrast to the general two-parameter family solutions. Of course, this does not make sense in the derivation of the first law. Thus, one has to relax the horizon condition for $A_r$ but conversely this in general invalids the Wald entropy formula. For more discussions, we refer the readers to section \ref{vectortensor} and section \ref{nonmini1}.

The paper is organized as follows. In section 2, we study a certain type vector-tensor theories. We analyze the structure of the general static maximally symmetric solutions. We also briefly review the Wald formalism, derive explicit formulas for our gravity model and discuss the subtleties in the derivation of the first law. In section 3, we study the minimally coupled theory by introducing a quartic self-interaction term for the vector. We obtain the general static solution with $A_r\neq 0$ and study various properties of the solutions. From section 4 to section 6, we study non-minimally coupled vector-tensor theories and obtain a lot of exact black hole solutions depending on the parameters of the theories. We also derive the first law using Wald formalism. We conclude in section 7.

\section{The vector-tensor theories}\label{vectortensor}

\subsection{Structure of general static solutions}
In this paper, we consider Einstein gravity either minimally or non-minimally coupled to a vector field together with a potential $V$ (it should not be confused with the vector field $A$). The Lagrangian density is given by
\be\label{genela}
\mathcal{L}= R-\fft{1}{4}F^2-\beta A^2 R+\gamma R_{\mu\nu}A^\mu A^\nu-V(\psi) \,,
\ee
where $F=dA$ and $\psi\equiv A_{\mu}A^{\mu}$. Note that the effective gravitational coupling constant is inversely proportional to
\be \kappa_{\mathrm{eff}}=1-\beta A^2\,.\ee
To avoid ghost-like graviton modes, we require $\kappa_{\mathrm{eff}}$ being positive definite throughout this paper. In addition, the $\gamma$ coupling term can be written more explicitly as
\be R_{\mu\nu}A^\mu A^\nu=A^\mu\big(\nabla_\mu\nabla_\nu-\nabla_\nu\nabla_\mu \big)A^\nu \,.\ee
This is a special case discussed in \cite{Heisenberg:2014rta}, where a general construction of vector-tensor theories preserving parity has been well studied.

The covariant equations of motions are
\be\label{eom}
 G_{\mu\nu}=T_{\mu\nu}^{\rm (min)}+\beta Y_{\mu\nu}+\gamma Z_{\mu\nu}\,,\qquad
\nabla_\mu F^{\mu\nu} = 2A^\nu\Big(\beta R + \fft{dV}{d\psi}\Big)-2\gamma R^{\mu\nu}A_\mu\,,
\ee
where $ G_{\mu\nu}=R_{\mu\nu}-\fft 12 R g_{\mu\nu}$ is the Einstein tensor and
\bea
&&T_{\mu\nu}^{\rm (min)}=\fft{1}{2}\Big(F^2_{\mu\nu}-\fft 14 g_{\mu\nu}F^2  \Big)+\Big(\fft{dV}{d\psi}A_\mu A_\nu-\fft 12 g_{\mu\nu}V(\psi)\Big)\,,\nn\\
&&Y_{\mu\nu}= A^2 G_{\mu\nu}+\big(g_{\mu\nu}\Box-\nabla_\mu\nabla_\nu\big)A^2+ R A_\mu A_\nu\,,\nn\\
&&Z_{\mu\nu}=-2A^\sigma R_{\sigma(\mu}A_{\nu)}+\nabla_\sigma \nabla_{(\mu}\big (A_{\nu)}A^\sigma \big)\nn\\
&&\qquad\quad -\ft 12 \square \big( A_\mu A_\nu \big)+\ft 12\Big(R_{\alpha\beta}A^\alpha A^\beta-\nabla_\alpha \nabla_\beta \big( A^\alpha A^\beta \big) \Big)\,.
\eea
For later convenience, we denote the Einstein and the vector equations of motions in (\ref{eom}) by $E_{\mu\nu}=0$ and $\mathcal{P}^\mu=0$ respectively.

In particular, we are interested in a vector potential of the type
\be V=2\Lambda_0+\fft 12 m^2 A^2+\gamma_4 A^4 \,,\label{potential}\ee
where $\gamma_4$ is a coupling constant characterizing the self-interaction of the vector field. Hence, the general theories are characterized by five independent parameters $(\beta\,,\gamma\,,\Lambda_0\,,m^2\,,\gamma_4)$. For $\gamma=0$, the theories with such a potential were first studied in \cite{Fan:2016jnz} whilst the $\gamma_4=0$ case has been studied in \cite{Geng:2015kvs,
Chagoya:2016aar,Minamitsuji:2016ydr,Babichev:2017rti} for certain coupling constants but most of them are limited to the four dimension. Instead, in this paper we will investigate the theories for general coupling constants and solve the static maximally symmetric solutions in general dimensions.

The most simple solutions of the theories (\ref{genela}) are given by
\be G_{\mu\nu}=-\Lambda_0 g_{\mu\nu}\,,\qquad A=0 \,,\ee
It follows that depending on the sign of the bare cosmological constant, the maximally symmetric vacuum is AdS ($\Lambda_0<0$), Minkowski ($\Lambda_0=0$) or
dS ($\Lambda_0>0$) space-times, respectively. Linearizing the equations of motions around the vacuum, we find that the linear fluctuations of the
equations are described by a massless graviton and a Proca which has an effective mass
\be m^2_{\mathrm{eff}}=m^2+\ft{4\Lambda_0}{n-2}\big(n \beta-\gamma\big) \,,\ee
where $n$ denotes the space-time dimensions. Notice that owing to the existence of the non-minimal couplings, an effective Proca mass can be generated in the vacuum even if the bare mass vanishes. Likewise, even if the bare mass is nonzero, the U(1) gauge symmetry of the vector can be restored at the linear level when the parameters are such that $m^2_{\mathrm{eff}}=0$. This is true for any Ricci-flat metric, including Schwarzschild and Kerr black holes.

The most general ansatz for static maximally symmetric solutions is
\be\label{solansatz}ds^2=-h\,dt^2+\fft{dr^2}{f}+r^2 d\Omega_{n-2\,,k}^2\,,\qquad A=A_tdt+A_r dr  \,,\ee
where $h\,,f\,,A_t\,,A_r$ are all functions of $r$ and $d\Omega_{n-2\,,k}^2$ is the metric of the codimension-2 space with spherical/hyperbolic/toric symmetries, corresponding to $k=1\,,-1\,,0$, respectively. It is easy to see that the vector equation $\mathcal{P}^r$ is purely algebraic for $A_r$. We find
\be A_r\,\big(\gamma_4\, A_r^2+\cdots \big)=0 \,,\label{arcontraint}\ee
where the dotted term is composed of the functions $h\,,f\,,A_t$ and their derivatives with respect to $r$ ( this term exactly vanishes for Einstein-Proca theory and hence the solutions with a nonzero $A_r$ do not exist in this case. ). It is clear that the above equation has isolated roots $A_r=0$ and $A_r\neq 0$, corresponding to different branch solutions. Consequently, in general the solutions with $A_r\neq 0$ do not have a smooth limit to send $A_r\rightarrow 0$ and reduce to the solutions with $A_r=0$. This is also true even if $\gamma_4=0$, in which case $A_r$ in general can not be solved algebraically\footnote{ In fact, the Einstein equation $E_{rr}$ is also an algebraic equation for $A_r$ when $\beta=\gamma/2$. Thus, in this case, $A_r$ can still be solved algebraically even if $\gamma_4=0$.}. Nonetheless, the ansatz (\ref{solansatz}) is most general for both $A_r=0$ and $A_r\neq 0$ solutions. We will study either of the two cases or both of them, depending on whether we can solve exact black hole solutions.

In the near horizon region, the metric functions and the vector fields can be expanded as Taylor series of the form
\bea
&&h=(r-r_0)+h_2(r-r_0)^2+h_3(r-r_0)^3+\cdots\,,\nn\\
&&f=f_1(r-r_0)+f_2(r-r_0)^2+f_3(r-r_0)^3+\cdots\,,\nn\\
&&A_t=a_0+a_1(r-r_0)+a_2(r-r_0)^2+a_3(r-r_0)^3+\cdots\,,\nn\\
&&A_r=\ft{b_0}{(r-r_0 )^{\sigma}}\Big(1+b_1(r-r_0)+b_2(r-r_0)^2+b_3(r-r_0)^3+\cdots \Big)\,,
\label{horizon}\eea
where $r_0$ denotes the horizon radii and we have set $h_1=1$ owing to the scaling symmetry of the time coordinate. It should be emphasized that unlike the $A_r=0$ case, for the solutions with a non-vanishing $A_r$ the finite norm condition of the vector is insufficient to govern $A_t$ vanishes on the horizon. We find that $\sigma=1$ when $a_0\neq 0$ and $\sigma=1/2$ when $a_0=0$. Both cases are allowed by the equations of motions. Substituting the expansions into the equations of motions, we find that for the minimal theory $\beta=0=\gamma$ and a certain non-minimal theory with $\beta=\gamma/2$, there are either three independent parameters $(r_0\,,a_0\,,a_1)$ when $a_0\neq 0$ or two parameters $(r_0\,,a_1)$ when $a_0=0$ on the horizon. In these two cases the coefficient $f_i\,,b_i$ are completely fixed because the metric function $f$ and the vector field $A_r$ are solved algebraically from the equations $\mathcal{P}^r\,,E_{rr}$. For generic case, the near horizon solutions are characterized by four independent parameters: $(r_0\,,f_1\,,a_0\,,b_1)$ when $a_0\neq 0$ and $(r_0\,,f_1\,,a_1\,,b_0)$ when $a_0=0$. For all these cases, the rest of the coefficients can be solved in terms of functions of the two, three or four independent parameters.

 However, in spite of that a nonzero $a_0$ is compatible with the equations of motions, it leads to a divergent local diffeomorphism invariant of the vector $A_{\bar{a}}=E_{\bar a}^{\mu}A_{\mu}$ on the horizon, where $E^\mu_{\bar a}$ is the inverse vielbein. This is something that we do not appreciate\footnote{One of the disasters of a divergent $A_{\bar a}$ is it leads to a divergent Wald formula $\delta H$ on the horizon. } and we will not discuss this case further in the remaining of this paper.

The general structure of the asymptotic solutions at infinity heavily depends on the five parameters of the theories as well as the asymptotical structure of the space-times. Here we shall not analyze them in a case-by-case basis since most of the solutions that we obtain contain all the independent integration constants. Nevertheless, it is deserved to show some universal aspects of the general asymptotic solutions. We find
\be
h=\cdots+g^2r^2+k_{\mathrm{eff}}-\fft{2\mu}{r^{n-3}}+\cdots\,,\quad  A_t=\cdots+q_1-\fft{q_2}{r^{n-3}}+\cdots\,,
\ee
where the effective cosmological constant is parameterized by $\Lambda_{\mathrm{eff}}=-\ft 12(n-1)(n-2)g^2$, $k_{\mathrm{eff}}$ is a function of $(k\,,\mu\,,q_1\,,q_2)$ and in general $k_{\mathrm{eff}}\neq k$ (we call it the effective curvature of the codimension-2 space). It is clear that the asymptotic solutions are characterized by three independent integration constants $\mu\,,q_1\,,q_2$  which are associated with the black hole mass and the vector charges respectively. However, only two of the three parameters are truly independent since the boundary conditions on the horizon provide an algebraic constraint for the three parameters. For example, we may take the parametric relation by saying $q_1=q_1(\mu\,,q_2)$. Then the full solutions are characterized by two independent parameters $\mu\,,q_2$, which are analogous to the case of a Reissner-Nordstr\o m (RN) black hole.

\subsection{ Wald formalism and thermodynamics}\label{waldthermo}
 In this paper, we will adopt the Wald formalism to derive the first law of thermodynamics for all the solutions we obtain. The Wald formalism provides a systematic procedure for the derivation of first law of thermodynamics for the solutions of a generic gravity theory. It was first developed by Wald in \cite{wald1,wald2}. Variation of the action with respect to the metric and the matter fields, one finds
\be \delta \Big(\sqrt{-g}\mathcal{L} \Big)=\sqrt{-g}\Big(E_\phi \delta \phi+\nabla_\mu J^{\mu}\Big) \,,\ee
where $\phi$ collectively denotes the dynamical fields and $E_\phi=0$ are the equations of motions. For our gravity model, the current $J^\mu$ receives
contributions from both the gravity and the vector.
We find
\bea
&&J^\mu=J_{(G)}^\mu+J_{(A)}^\mu\,,\qquad J^\mu_{(G)}=G^{\mu\nu\rho\sigma}\nabla_\nu \delta g_{\rho\sigma}    \,,\nn\\
&&J^\mu_{(A)}=-F^{\mu\nu}\, \delta A_\nu+\beta\, G^{\mu\nu\rho\sigma}\big(\nabla_\nu A^2-A^2\nabla_\nu \big)\delta g_{\rho\sigma}+\gamma\, J^\mu_{(\gamma)}\,,
\eea
where $G^{\mu\nu\rho\sigma}$ is the Wheeler-Dewitt metric, defined by
\be G^{\mu\nu\rho\sigma}=\frac 12 (g^{\mu\rho}g^{\nu\sigma}+g^{\mu\sigma}g^{\nu\rho})-g^{\mu\nu}g^{\rho\sigma}\,,\ee
and the current associated with the $\gamma$ coupling term is
\bea
&&J^\mu_{(\gamma)}=\Big( g^{\mu\lambda}A^\rho A^\sigma \nabla_\sigma \delta g_{\lambda\rho}-\nabla^\lambda \big( A^\rho A^\mu \big)\delta g_{\lambda\rho}\Big)+\fft 12\Big( \nabla^\mu\big( A^\lambda A^\rho \big)\delta g_{\lambda\rho}\nn\\
&&\qquad \quad-A^\lambda A^\rho \nabla^\nu \delta g_{\lambda\rho}  \Big)
+\fft 12 g^{\lambda\rho} \Big( \nabla_\sigma\big(A^\mu A^\sigma \big)\delta g_{\lambda\rho}-A^\mu A^\sigma \nabla_\sigma \delta g_{\lambda\rho}   \Big)\,.
\eea
Note that we have put the current associated with the non-minimally coupled terms into the vector sector. For a given current $J^\mu$,  one can define a current 1-form and its Hodge dual as
\be
J_\1=J_\mu dx^\mu\,,\qquad
\Theta_{\sst{(n-1)}}={*J_\1}\,.
\ee
When the variation is generated by an infinitesimal diffeomorphism $\xi^\mu=\delta x^\mu$, one can define an associated Noether current $(n-1)$-form as
\be
J_{\sst{(n-1)}}=\Theta_{\sst{(n-1)}}-i_\xi\cdot {*\mathcal{L}}\,,
\ee
where $i_\xi\cdot$ denotes the contraction of $\xi$ with the first index of the $n$-form ${}^*\mathcal{L}$ it acted upon. It was shown in \cite{wald1,wald2} that the Noether current $J_{(n-1)}$ is closed once the equations of motions are satisfied, namely
\be dJ_{(n-1)}=\mathrm{e.o.m} \,,\ee
where e.o.m denotes the terms proportional to the equations of motions. Thus one can further define a charge $(n-2)$-form as
\be J_{(n-1)}=dQ_{(n-2)} \,.\ee
It was shown in \cite{wald1,wald2} that when $\xi$ is a Killing vector, the variation of the Hamiltonian with respect to the integration constants of a specific solution is
given by
\be
\delta H=\frac{1}{16\pi}\Big[\delta \int_{\mathcal{C}}J_{\sst{(n-1})}- \int_{\mathcal{C}}d(i_\xi\cdot \Theta_{\sst{(n-2)}})\Big]=\frac{1}{16\pi}\int_{\Sigma_{n-2}}\Big[\delta Q_{\sst{(n-2)}}-i_\xi\cdot \Theta_{\sst{(n-2)}}\Big]\,.\label{generalwald}
\ee
where $\mathcal{C}$ is a Cauchy surface, $\Sigma_{n-2}$ is its two boundaries, one on the horizon and the other at infinity. For our vector-tensor theories,
it is straightforward to derive the various quantities in the Wald formalism though the calculations are a little lengthy. For pure gravity, we have \cite{wald2}
\bea
J_{\sst{(n-1)}}^{(G)} &=&-2\varepsilon_{\mu c_1...c_{n-1}} \nabla_\nu\Big(\nabla^{[\mu}\xi^{\nu]}
\Big)  \,,\cr
Q_{\sst{(n-2)}}^{(G)} &=& -\varepsilon_{\mu\nu c_1...c_{n-2}}\,\nabla^{\mu}\xi^{\nu}\,,\cr
i_{\xi}\cdot \Theta_{(n-1)}^{(G)}&=&\varepsilon_{\mu\nu c_1...c_{n-2}}\xi^{\nu}\Big(
G^{\mu\lambda\rho\sigma} \nabla_\lambda \delta g_{\rho\sigma}\Big)\,.\label{various}
\eea
For the vector sector, we obtain
\bea
&&J^{(A)}_{(n-1)}=2\varepsilon_{\mu c_1\cdots c_{n-1}}\nabla_\nu\Big[-\ft 12 F^{\mu\nu}A^\sigma\xi_\sigma+\beta \Big(A^2 \nabla^{[\mu}\xi^{\nu]}+2\xi^{[\mu}\nabla^{\nu]}A^2 \Big)\nn\\
&&\qquad \qquad \qquad \qquad\qquad +\gamma\Big( \xi_\sigma \nabla^{[\mu}\big( A^{\nu]}A^\sigma\big)
-\xi^{[\mu}\nabla_\sigma\big( A^{\nu]}A^\sigma \big)-A^\sigma A^{[\mu}\nabla_\sigma \xi^{\nu]} \Big)\Big] \,,\nn\\
&&  Q_{(n-2)}^{(A)}=\varepsilon_{\mu\nu c_1\cdots c_{n-2}}\Big[-\ft 12 F^{\mu\nu}A^\sigma\xi_\sigma+\beta \Big(A^2 \nabla^{\mu}\xi^\nu+2\xi^\mu \nabla^\nu A^2\Big)\nn\\
&&\qquad\qquad\qquad \qquad\quad +\gamma\Big( \xi_\sigma \nabla^{\mu}\big( A^{\nu}A^\sigma\big)
-\xi^{\mu}\nabla_\sigma\big( A^{\nu}A^\sigma \big)-A^\sigma A^{\mu}\nabla_\sigma \xi^{\nu} \Big) \Big]\,, \nn\\
&& i_\xi\cdot \Theta_{(n-1)}^{(A)}=\varepsilon_{\mu\nu c_1\cdots c_{n-2}}\xi^\nu\Big(-F^{\mu\nu}\, \delta A_\nu
+\beta\, G^{\mu\nu\rho\sigma}\big(\nabla_\nu A^2-A^2\nabla_\nu \big)\delta g_{\rho\sigma}+\gamma J^\mu_{(\gamma)} \Big) \,.
\eea
Notice that the Wald formalism does not explicitly depend on the non-derivative terms of the Lagrangian density. The various quantities have been given in \cite{Fan:2016jnz} for $\gamma=0$ and in \cite{Liu:2014tra,Liu:2014dva,Fan:2014ixa,Fan:2014ala,Fan:2015yza,Chen:2016qks} for $\beta=0=\gamma$.

Now we evaluate $\delta H$ for the general static solutions with maximal symmetry (\ref{solansatz}). Let $\xi=\partial/\partial t$, we obtain
\bea\label{waldformula1}
&&\delta H=\delta H^{(G)}+\delta H^{(A)}\,\nn\\
&&\delta H^{(G)}= \fft{\omega_{n-2}}{16\pi} r^{n-2}\, \sqrt{\fft{h}{f}}\, \Big(-\fft{n-2}{r} \Big)\delta f\,,
\eea
and
\bea\label{waldformula2}
&&\delta H^{(A)}=\delta H^{(A)}_{(\mathrm{min})}+\delta H^{(\beta)}_{(\mathrm{non})}+\delta H^{(\gamma)}_{(\mathrm{non})}\,,\nn\\
&&\delta H^{(A)}_{(\mathrm{min})}=-\frac{\omega_{n-2}}{16\pi}r^{n-2}\sqrt{\frac{h}{f}}\, \Big(\frac{f}{h}A_t\delta A_t'+\frac 12 A_t A_t'\big(\frac{\delta f}{h}-\frac{f\delta h}{h^2}\big)\Big)\,,\nn\\
&&\delta H^{(\beta)}_{(\mathrm{non})}=\fft{\beta\,\omega_{n-2}}{16\pi}r^{n-2}\sqrt{\fft{f}{h}}\Big(\fft{6h'}{h}A_t\delta A_t-4\delta(A_t A'_t)+A_t^2\, \Delta_1+ h f A_r^2\,\Delta_2\Big)\,,\nn\\
&&\delta H^{(\gamma)}_{(\mathrm{non})}=-\fft{\gamma\,\omega_{n-2}}{16\pi}r^{n-2}\sqrt{\fft{f}{h}}\Big(\fft{3h'}{h}A_t\delta A_t-2\delta(A_t A'_t )+A_t^2\, \Sigma_1+h f A_r^2\, \Sigma_2 \Big)\,.
\eea
where $\omega_{n-2}$ is the volume factor of the $(n-2)$ dimensional space and
\bea\label{waldformula3}
&& \Delta_1=\fft{2\delta h'}{h}+\Big(\fft{4A_t'}{A_t}-\fft{5h'}{h}\Big)\fft{\delta h}{h}-\Big(\fft{2A_t'}{A_t}-\fft{h'}{h}+\fft{n-2}{r} \Big)\fft{\delta f}{f}\,,\\
&& \Delta_2=\fft{4\delta A'_r}{A_r}+\Big( \fft{2A'_r}{A_r}-\fft{h'}{h}+\fft{2f'}{f} \Big)\fft{2\delta A_r}{A_r}+\fft{2\delta f'}{f}
+\Big(\fft{6A'_r}{A_r}-\fft{h'}{h}+\fft{f'}{f}+\fft{n-2}{r} \Big)\fft{\delta f}{f}\,,\nn\\
&& \Sigma_1=\fft{\delta h'}{h}+\Big(\fft{2A_t'}{A_t}-\fft{5h'}{2h}\Big)\fft{\delta h}{h}-\Big(\fft{A_t'}{A_t}-\fft{h'}{2h} \Big)\fft{\delta f}{f}\,,\nn\\
&& \Sigma_2=\fft{2\delta A'_r}{A_r}+\Big( \fft{2A'_r}{A_r}-\fft{h'}{h}+\fft{2f'}{f}+\fft{2(n-2)}{r} \Big)\fft{\delta A_r}{A_r}+\fft{\delta f'}{f}
+\Big(\fft{3A'_r}{A_r}-\fft{h'}{2h}+\fft{f'}{2f}+\fft{2(n-2)}{r} \Big)\fft{\delta f}{f}\,.\nn
\eea
It was shown in \cite{wald1,wald2} that evaluating $\delta H$ on the horizon yields
\be  \delta H_+=T \delta S \,,\ee
where the temperature and Wald entropy are given by
\be
T=\fft{\kappa}{2\pi}\,,\qquad S=-\fft{1}{8} \int_+ \sqrt{h}\, d^{n-2}x\, \epsilon^{ab}\epsilon^{cd}\, \fft{\partial L}{\partial R^{abcd}}\,.
\ee
Here $\kappa$ is the surface gravity on the horizon. Throughout this paper, the Wald entropy is always denoted by $S$, without any subscript. For our metric ansatz, we have
\be T=\fft{1}{4\pi}\sqrt{h'(r_0)f'(r_0)}\,,\qquad S=\ft 14\mathcal{A}\Big[ 1+(\beta+\ft 12\gamma)\Big( \ft{A^2_t(r_0)}{h(r_0)}- A_r^2(r_0)f(r_0)\Big) \Big] \,,\label{entropyexp}\ee
where $\mathcal{A}=\omega_{n-2} r_0^{n-2}$ is the area of the horizon. Evaluating $\delta H$ at both infinity and on the event horizon yields
\be \delta H_\infty=\delta H_+ \,.\ee
Thus the first law of thermodynamics is simply
\be \delta H_\infty=T \delta S \,.\label{waldeq}\ee
This is the standard derivation of the first law when the Wald entropy formula holds. However, the situation in our case is even more subtle because counterintuitively, the finite norm condition of the vector is not sufficient to govern the validity of the Wald entropy formula (\ref{entropyexp}). The reason is $\delta H_+$ may be non-integrable for general near horizon solutions. As was discussed in \cite{Feng:2015oea}, to govern the validity of Wald entropy formula one should require the local diffeomorphism invariant $A_{\bar a}$ of the vector vanishes on the horizon\footnote{ The Wald entropy is closely related to the Noether charge as: $\fft{1}{16\pi}\int_{r=r_0} Q_{(n-2)}=T\,S $. So the variation of Hamiltonian on the horizon is
\be \delta H_+=T \delta S+\Big(S \delta T-\fft{1}{16\pi}\int_{r=r_0}i_\xi\cdot \Theta_{(n-1)}  \Big) \,. \nn\ee
Here the cancellation of the second term on the r.h.s of this equation requires $A_{\bar a}$ vanishes.}. However, for our vector-tensor theories such a condition in general turns out to be too strong to be imposed because $A_r$ is an algebraic degree of freedom which does not have corresponding vector charges. So we have to relax the condition for $A_{\bar r}(r_0)$ and simply demand a vanishing $A_{\bar t}(r_0)$. This has fixed the parametric relation between the parameters $(\mu\,,q_1\,,q_2)$ of the asymptotic solutions but it does not necessarily lead to a vanishing $A_{\bar r}(r_0)$. Consequently, $\delta H_+$ in general becomes non-integrable. We find
\bea\label{waldhorizon}
\delta H_+&=&\ft 14 T \Big[ \Big(1-(\beta-\gamma)\Phi \Big)\delta \mathcal{A}-(\beta-\ft 12 \gamma)\mathcal{A}\,\delta \Phi   \Big]    \nn\\
&=&T \Big[ \delta S+\ft 14 \gamma \Big(\mathcal{A}\, \delta \Phi+\ft 32 \Phi\, \delta\mathcal{A} \Big)  \Big] \,.
\eea
where $\Phi\equiv A^2_{\bar r}(r_0)$ is a dimensionless quantity. The existence of the non-integrable one-form on the r.h.s of the equation invalids the Wald entropy formula as well as a refining entropy defined as $\delta H_+\equiv T\delta S_{\mathrm{re}}$. Nonetheless, formally one can still write down a ``first law" using the Wald equation despite that its physical meaning is unclear. Notice that when $\gamma=0$, one will not encounter the trouble any longer because of $\delta H_+=T\delta S$. Furthermore, when $\beta=\ft 12 \gamma$, $\delta H_+$ is integrable as well\footnote{For $\beta=\gamma/2$, the non-minimal coupling term becomes $G_{\mu\nu}A^\mu A^\nu$. In this case, the solutions with  $A_\mu=\partial_\mu \phi$ are connected to those of Horndeski gravity. However, the dynamics of the two theories are significantly different. It was shown in \cite{Feng:2015oea} that in Horndeski gravity $\delta H_+$ is always integrable but $\delta H_+=T \delta \bar S\neq T\delta S$. This is very different from our results.}, given by
 \be \delta H_+=\ft 14 T \big( 1+\ft 12\gamma\Phi \big)\delta \mathcal{A}\equiv \widetilde{T}d\widetilde{S} \,,\label{nonhorizon2}\ee
where the {\it improved temperature} and entropy are defined by
\be \widetilde{T}\equiv \big(1+\ft 12\gamma \Phi \big)T\,,\qquad \widetilde{S}\equiv \big(1-\gamma\Phi \big)^{-1}S=\ft 14\mathcal{A}  \,.\label{improvedtementropy}\ee
Here comes an intriguing question that how the {\it improved temperature} $\widetilde{T}$ is interpreted in the thermodynamical content. We leave this as a future direction for research.
For generic case, as will be shown later, the above non-integrable one-form may vanish for a certain coupling constant $\gamma$.

\section{Minimal theory}

In this section, we study a minimally coupled theory described by
\be\label{simla1}
\mathcal{L}= R-2\Lambda_0-\fft{1}{4}F^2-\fft{1}{2}m^2 A^2-\gamma_4 A^4\,,
\ee
which generalizes the Einstein-Proca theory with a quartic self-interaction term for the vector field. Despite the simple form of the theory, there are some new interesting and important features in the theory. For instance, although for $\Lambda_0 \neq 0$ the maximally symmetric vacuum of the theory is (A)dS space-times, it also allows a simple solution which is Minkowski space-times supported by a constant vector
\be\label{minisimsol}
ds^2=-dt^2+dr^2+r^2 d\Omega_{n-2}^2\,,\qquad A=q_1 dt+\sqrt{q_1^2-\ft{m^2}{4\gamma_4}}\, dr\,,
\ee
provided the parametric relation
\be \Lambda_0=\fft{m^4}{32\gamma_4 } \,.\label{crit}\ee
Note that this relation leads to a perfect squared vector potential $V=-\gamma_4\big(A^2+\ft{m^2}{4\gamma_4} \big)^2$ and the parameters in the solution (\ref{minisimsol}) are such that $V=0$. Reality of the solution naturally requires $q_1^2\geq \ft{m^2}{4\gamma_4}$, where the ``$=$" case corresponds to a vanishing $A_r$, which was first studied in \cite{Fan:2016jnz}. It is worth emphasizing that the above solution (\ref{minisimsol}) is not a vacuum solution because the vector breaks the gauge symmetries explicitly. When the bare cosmological constant deviates from the critical value (\ref{crit}), an effective cosmological constant emerges
\be \Lambda_{\mathrm{eff}}=\Lambda_0-\ft{m^4}{32\gamma_4}\,,\label{minicosm}\ee
in the corresponding solutions because the potential now becomes
 \be V=2\Lambda_{\mathrm{eff}}-\gamma_4\big(A^2+\ft{m^2}{4\gamma_4} \big)^2 \,.\ee

A second new and probably more important feature of the theory (\ref{simla1}) is that we can exactly solve its general static maximally symmetric black hole solution with $A_r\neq 0$. This is quite surprising since up to now any exact black hole solution has not been found in Einstein-Proca theory. To keep generality, let's discuss how to analytically solve the equations of motions for general parameters.

First, the equations $\mathcal{P}^r$ and $E_{rr}$ are purely algebraic for $A_r$ and $f$ so we can solve the two functions in terms of $h\,,A_t$ and their derivatives
\be\label{minialgebraic}
f=\fft{4\gamma_4 A_t^2-m^2 h}{4\gamma_4 h A_r^2}\,,\quad
A_r^2=\ft{2\big(4\gamma_4 A_t^2-m^2 h \big)\Big(r^2 A_t^{'^2}+2(n-2)r h'+2(n-2)(n-3)h \Big)}{\Big((m^4-32\gamma_4 \Lambda_0)r^2+16(n-2)(n-3)\gamma_4 k \Big)h^2}\,.
\ee
The remaining independent equations are $E_{tt}$ and $\mathcal{P}^t$, which are second order non-linear ordinary differential equations (ODE) of $A_t$ and $h$. They are in general very difficult to integrate. Fortunately, we find that the two equations become integrable if we parameterize the two functions as
\be\label{minipara} h=-\fft{2\mu}{r^{n-3}}+\fft {1}{4r^{n-3}}\int \mathrm{d}r\,H(r)\,,\quad A_t=\int\mathrm{d}r \, r^{-\fft{(n-2)}{2}}\sqrt{F(r)-\ft{n-2}{2}H(r)}\,,\ee
where $\mu$ is an integration constant. Strikingly, the equation $E_{tt}$ simplifies to a single linear first order ODE for $F$
\bea
0&=&\Big((m^4-32\gamma_4\Lambda_0)r^2+16(n-2)(n-3)\gamma_4 k \Big)F'\nn\\
&&-\ft{(n-2)}{r}\Big( (m^4-32\gamma_4\Lambda_0)r^2+16(n-3)(n-4)\gamma_4 k \Big)F \,,
\eea
which can be immediately solved by
\be F=\fft{C}{8\gamma_4}\Big((m^4-32\gamma_4\Lambda_0)r^2+16(n-2)(n-3)\gamma_4 k \Big)r^{n-4} \,,\label{miniF}\ee
where $C\neq 0$ is a new integration constant. Substituting (\ref{miniF}) into $\mathcal{P}^r$, we find it also reduces to a linear first order ODE
\be 0=H'+\fft{n-2}{r}H-\fft{C}{2\gamma_4}\Big((m^4-32\gamma_4\Lambda_0)r^2+16(n-3)^2\gamma_4 k \Big)r^{n-5} \,,\ee
which is easy to integrate. We get
\be H=\fft{\big(m^4-32\gamma_4\Lambda_0 \big) C}{4(n-2)\gamma_4}r^{n-2}+4(n-3)C k\, r^{n-4}-\fft{2(n-3)^2 q_2^2}{(n-2)r^{n-2}} \,,\label{miniH}\ee
where $q_2$ is an integration constant associated with the Coulomb-like charge of the vector, as will be shown later. Plugging Eq.(\ref{miniF}) and Eq.(\ref{miniH}) into (\ref{minipara}), we obtain
\be h=C\big(g^2 r^2+k \big)-\fft{2\mu}{r^{n-3}}+\fft{(n-3)q_2^2}{2(n-2)r^{2n-6}} \,,\qquad A_t=q_1-\fft{q_2}{r^{n-3}} \,,\label{minihAt}\ee
where the effective cosmological constant (\ref{minicosm}) is parameterized by $\Lambda_{\mathrm{eff}}=-\ft 12(n-1)(n-2)g^2$ and $q_{1,2}$ are the two vector charges which are analog of the chemical potential/charge density of Reissner-Nordstr\o m (RN) black hole. Now it is clear that $C$ is a non-physical parameter which is associated with the scaling symmetry of the time coordinate. Without loss of generality, we set $C=1$. Finally, substituting (\ref{minihAt}) into (\ref{minialgebraic}), we find
\be f=h \,,\qquad  A_r^2=\fft{A_t^2}{f^2}-\fft{m^2}{4\gamma_4 f} \,.\ee
This completes our derivation. To conclude, we obtain
\bea && ds^2=-fdt^2+\fft{dr^2}{f}+r^2d\Omega_{n-2\,,k}^2\,,\quad  A_r=\sqrt{\ft{A_t^2}{f^2}-\ft{m^2}{4\gamma_4 f}}\,,\nn\\
     && A_t=q_1-\fft{q_2}{r^{n-3}}\,,\quad f=g^2r^2+k-\fft{2\mu}{r^{n-3}}+\fft{(n-3)q_2^2}{2(n-2)r^{2n-6}} \,.
\label{sol1ads}\eea
Now we are ready to give some comments on the solution. First, as is clear from the derivation, the above solution is the most general static solution with maximal symmetries in the presence of a non-vanishing $A_r$. It contains all the three integration constants $\mu\,,q_1\,,q_2$ which are associated with the black hole mass and the vector charges respectively. Second, formally without $A_r$ the solution is simply the RN black hole! In fact, this is easily understood because under the special ansatz $h=f\,,A_r=\sqrt{\ft{A_t^2}{f^2}-\ft{m^2}{4\gamma_4 f}}$, the norm of the vector is a constant $A^2=-\ft{m^2}{4\gamma_4}$ such that $V=2\Lambda_{\mathrm{eff}}$ and the stress tensor of the vector reduces to that of a Maxwell: $T^{A}_{\mu\nu}=T^{\mathrm{Max}}_{\mu\nu}$. In this sense, we may call the solution a {\it stealth Reissner-Nordstr\o m black hole} (but one should remember the gauge symmetry is breaking).

Moreover, evaluating $\delta H$ at infinity yields
\be \delta H_\infty=\delta M-\fft{(n-3)\,\omega_{n-2}}{16\pi} q_1\delta q_2 \,,\ee
where the black hole mass $M$ is defined by\footnote{In this paper, without specification, we always define the mass using the usual fall-off mode $1/r^{n-3}$ associated with the condensate of the massless gravitons. It is the standard ADM/AMD mass of the asymptotically flat/AdS solutions for the minimal theory. However, for non-minimal theory the situation is not so simple. Nonetheless, we continue using the concept for all these theories. }
\be M=\fft{(n-2)\,\omega_{n-2}}{8\pi}\,\mu \,.\label{mass}\ee
The Wald equation (\ref{waldeq}) implies the first law is
\be dM=T dS+\fft{(n-3)\,\omega_{n-2}}{16\pi} q_1  dq_2\,,\label{minifirstlaw}\ee
where the temperature and entropy are given by
\be T=\fft{1}{4\pi r_0}\Big((n-1)g^2r_0^2+(n-3)k-\ft{(n-3)^2q_2^2}{2(n-2)r_0^{2n-6}} \Big)\,,\qquad S=\ft 14 \mathcal{A}   \,.\ee
In addition, the Smarr relation is
\be M=\fft{n-2}{n-3}T S+\fft{(n-3)\,\omega_{n-2}}{16\pi}q_1 q_2-\fft{2}{n-3}V P \,,\label{minismarr}\ee
where the thermodynamic pressure $P$ and volume $V$ are defined by
\be P=-\fft{\Lambda_{\mathrm{eff}}}{8\pi}\,,\qquad V=\fft{\omega_{n-2}}{n-1}r_0^{n-1} \,.\ee
All these global properties and the relations are exactly the same as the RN black hole in spite of that in general a global charge cannot be defined for the vector field due to the absence of a Gauss's law. By plugging the mass, temperature and entropy into the first law (\ref{minifirstlaw}), we ensure that the first law is valid if and only if $A_{\bar t}$ vanishes on the horizon. This is consistent with our previous argument. The result has nothing to do with $A_{\bar r}$ because the Wald formula Eq.(\ref{waldformula1}-\ref{waldformula3}) do not explicitly depend on $A_{r}$ for a minimal theory.

In particular, it is interesting to note that the third terms associated with the vector field on the r.h.s of Eq.(\ref{minifirstlaw}) and Eq.(\ref{minismarr}) look much like the electrostatic potential and electric charge of a RN black hole. Formally, we can introduce
\be \Phi_p\equiv q_1\,,\qquad  Q_p\equiv \fft{(n-3)\omega_{n-2}}{16\pi}\,q_2 \,,\ee
as the counterparts of the thermodynamic conjugate $(\Phi_e\,,Q_e)$ of a RN black hole. In the latter case, the non-integrable term $\Phi_e dQ_e$ associated to the Maxwell field in the first law is well understood as the working term of the electrostatic force. Thus, in this case the black hole mass is well defined via the first law by using the Wald entropy. Here comes an intriguing question: for our vector field whether the term $\Phi_p dQ_p$ appearing in the first law Eq.(\ref{minifirstlaw}) can be interpreted as the work of the force associated to the vector field as well. We find surprisingly, the answer is yes! The reason is for our solution the vector field enjoys a detailed balance condition $A^2=-m^2/(4\gamma_4)$ such that its self-interaction terms on the r.h.s of the equation of motion Eq.(\ref{eom}) are exactly cancelled. Thus, in this sense the vector charge $Q_p$ is globally conserved and the mass in addition to the ADM definition is well defined via the first law by making use of the Wald entropy. In addition, we also find that the absence of a naked curvature singularity at the origin leads to an upper bound for the vector charge
\be Q_p/M\leq \sqrt{\ft{n-3}{2(n-2)}} \,,\ee
where the bound is saturated for an extremal solution which is asymptotically flat.

Third, in general the limit $A_r\rightarrow 0$ is not allowed except for the asymptotically flat solution
\bea && ds^2=-fdt^2+\fft{dr^2}{f}+r^2d\Omega_{n-2}^2\,,\quad  A_r=\sqrt{\ft{A_t^2}{f^2}-\ft{m^2}{4\gamma_4 f}}\,,\nn\\
     && A_t=q_1-\fft{q_2}{r^{n-3}}\,,\quad f=1-\fft{2\mu}{r^{n-3}}+\fft{(n-3)q_2^2}{2(n-2)r^{2n-6}} \,.
\label{sol1}\eea
Note that the limit $A_r\rightarrow 0$ gives rise to an extremal solution \cite{Fan:2016jnz} because of $f\sim A_t^2$.

Finally, the reality of $A_r$ at any position of the space-times strongly constraints the parameters of the solution (here we do not clearly distinguish the integration constants of the solution and the coupling constants of the theory).  We find
\be \ft{m^2 g^2}{\gamma_4}\leq 0\,,\quad q_1^2\geq \ft{m^2}{4\gamma_4}\,k\,,\quad q_1q_2\leq \ft{m^2}{4\gamma_4}\, \mu \,,\quad \ft{m^2}{4\gamma_4}\leq \ft{2(n-2)}{n-3} \,.\ee
Note that the limit of a free vector field $\gamma_4\rightarrow 0$ is not well defined for our solutions.


\section{Non-minimal theory: case I}\label{nonmini1}
From now on, we turn to study the generalized Einstein-Proca theories with non-minimally coupled terms. In this section, we study a simple theory which has a single $\gamma$ term, namely
\be \mathcal{L}=R-\fft 14F^2+\gamma R_{\mu\nu}A^\mu A^\nu \,, \label{simla2}\ee
whilst the theory with a single $\beta$ term was studied in \cite{Fan:2016jnz}. It turns out that for this simple theory, we can find different kinds of stealth black hole solutions which satisfy $G_{\mu\nu}=0=T_{\mu\nu}$, depending on the non-minimal coupling constant.


\subsection{Stealth black hole: $\gamma=1$}

The first case we consider is when $\gamma=1$, we always have $T_{\mu\nu}=0$ for the special ansatz $h=f\,,A_t\propto f\,,A_r=0$. Therefore, we easily find a stealth Schwarzschild black hole solution
\bea\label{nonsol1}
&& ds^2=-fdt^2+\fft{dr^2}{f}+r^2 d\Omega^2_{n-2}\,,\nn\\
&& A=q_1 f dt\,,\qquad f=1-\fft{2\mu}{r^{n-3}}\,,
\eea
where $q_1\,,\mu$ are two independent integration constants associated with the vector charge and the black hole mass. The solution can be trivially generalized to including a cosmological constant, additional matter fields (such as a Maxwell field) or higher curvature terms (such as the Love-Lock terms ) in the Lagrangian density.

However, the solution is not most general since it contains one less integration constant. To derive its first law of thermodynamics, we shall first analyze the structure of general asymptotic solutions and derive the corresponding first law. We find
\bea
&&A_t=q_1-\fft{q_2}{r^{n-3}}+\cdots\,,\nn\\
&&h=1-\fft{2\mu}{r^{n-3}}-\fft{(n-3)q_2^2}{2(n-2)r^{2n-6}}+\cdots\,,\nn\\
&&f=1-\fft{2\mu}{r^{n-3}}-\fft{3(n-3)q_2^2}{2(n-2)r^{2n-6}}+\cdots \,.
\eea
It is easy to see that in general $h\neq f$ and the mass of the black hole does not receive contributions from the back-reaction of the vector. Substituting the asymptotic solutions into the Wald formula, we obtain
\be \delta H_\infty=\delta M+\fft{(n-3)\,\omega_{n-2}}{16\pi}\big(q_1\delta q_2+2q_2\delta q_1\big)-\fft{(n-3)\,\omega_{n-2}}{8\pi}\big(q_1^2\delta \mu+\ft 32 \mu\,\delta (q_1^2)\big) \,,\ee
where the mass is defined by (\ref{mass}). It follows that the first law reads
\be dM=T dS-\fft{(n-3)\,\omega_{n-2}}{16\pi}\big(q_1d q_2+2q_2  d q_1\big)+\fft{(n-3)\,\omega_{n-2}}{8\pi}\big(q_1^2 d \mu+\ft 32 \mu\,d (q_1^2)\big) \,.\label{nonfirstlaw1}\ee
It is interesting to note that there are two new pairs of thermodynamic conjugates: $(q_1\,,q_2)$ and $(\mu\,,q_1^2)$. This is very different from the Einstein-Proca black hole \cite{Liu:2014tra} which only has the first pair of conjugates. For the special solution (\ref{nonsol1}), we have $q_2=2\mu q_1$ and
\be T=\fft{n-3}{4\pi r_0}\,,\qquad S=\ft 14 \mathcal{A} \,.\ee
It is straightforward to verify the first law (\ref{nonfirstlaw1}) is indeed satisfied. In addition, define a new energy function as
\be E\equiv M+\fft{(n-3) \omega_{n-2}}{16\pi} q_1\big(q_2-2\mu q_1 \big) \,,\ee
the first law can be cast into the form of
\be dE=T dS-\fft{(n-3) \omega_{n-2}}{16\pi}\big(q_2-2\mu q_1 \big)d q_1 \,.\ee
So we may take $E$ as a function of $E=E(S\,,q_1)$. Note that the coupling constant $\gamma$ is dimensionless so the above equation contains all the dimensionful quantities in the theory and the solution. It follows that using the scaling dimensional arguments
\be E\rightarrow \lambda^{n-3}E\,,\qquad S\rightarrow \lambda^{n-2}S\,,\qquad q_1\rightarrow q_1 \,,\ee
we can derive a Smarr relation
\be E=\fft{n-2}{n-3}T S \,.\ee
Written back in terms of the original mass function, we find
\be M=\fft{n-2}{n-3}T S-\fft{(n-3) \omega_{n-2}}{16\pi} q_1\big(q_2-2\mu q_1 \big) \,.\ee
This can be easily verified although the solution (\ref{nonsol1}) is degenerate.

\subsection{Stealth black hole: generic case}\label{sec4.2}

\subsubsection{The solution and thermodynamics}
Interestingly, for generic $\gamma\neq 0$, we can also obtain a stealth Schwarzschild black hole solution which has a nonzero $A_r$. First, the vector equation $\mathcal{P}^r$ reads
\be 0=A_r\Big( \ft{h''}{h}-\ft{h'^2}{h^2}+\big(\ft{h'}{h}+\ft{2(n-2)}{r} \big)\ft{f'}{f} \Big) \,,\ee
while the Einstein equation $E_{rr}$ is no longer algebraic for $A_r$ or the metric functions. To proceed, we choose a special ansatz $h=f$. Then the vector equations dramatically simplify to
\be y''+\ft{n-2}{r}y=0\,,\qquad y=f\quad \mathrm{or}\quad y=A_t \,,\ee
which can be immediately solved as
\be h=f=1-\fft{2\mu}{r^{n-3}}\,,\qquad A_t=q_1-\fft{q_2}{r^{n-3}} \,.\ee
The remaining equations are $E_{tt}\,,E_{rr}$, both of which are ODE of $A_r$. Here, one may worry about the two equations are inconsistent with each other.
Fortunately, we are able to find an unique solution for $A_r$ which satisfies both equations
\bea\label{nonAr1}
A_r^2&=&\ft{r^{3-n}}{2(n-1)\gamma\mu f^2}\Big[2(n-3)(1-\gamma)\mu\, q_2^2\, r^{3-n}\nn\\
&&+\Big(2(n-2)\gamma-(n-3) \Big)q_2^2+4(n-1)\gamma\, \mu q_1(\mu q_1-q_2) \Big]\,.
\eea
This completes our derivation. In spite of that we do not expect to find the most general solutions of the theory at the very start, the solution we get contains all the three independent integration constants $\mu\,,q_{1\,,2}$. This gives us strong confidence that the above solution is the general static spherically symmetric black hole solution with a nonzero $A_r$. To govern the reality of $A_r$ at any position of the space-times, the coupling constant is bounded
\be 0<\gamma\leq 1 \,.\ee
For $\gamma=1$, we always have $A_r^2\geq 0$ so the parameters $\mu\,,q_1\,,q_2$ are free in this case. For $\gamma=\ft{(n-3)}{2(n-2)}$, the condition leads to $\mu q_1(\mu q_1-q_2)\geq 0$. For generic case, the constant terms in the square bracket of (\ref{nonAr1}) should be nonnegative.

For later convenience, we list the above solution as follows
\bea\label{nonsol2}
&& ds^2=-fdt^2+\fft{dr^2}{f}+r^2 d\Omega^2_{n-2}\,,\nn\\
&& A_t=q_1-\fft{q_2}{r^{n-3}}\,,\qquad f=1-\fft{2\mu}{r^{n-3}}\,,\nn\\
&&A_r^2=\ft{r^{3-n}}{2(n-1)\gamma\mu f^2}\Big[2(n-3)(1-\gamma)\mu\, q_2^2\, r^{3-n}\nn\\
&&\qquad +\Big(2(n-2)\gamma-(n-3) \Big)q_2^2+4(n-1)\gamma\, \mu q_1(\mu q_1-q_2) \Big]\,.
\eea
Evaluating $\delta H$ at infinity yields
\bea\label{nonHinfty}
\delta H_\infty&=&\delta \widetilde{M}+\fft{(n-2)\gamma\omega_{n-2}}{8\pi}\Big(q_2\delta q_1+\big(1-\ft{n-3}{2(n-2)\gamma} \big)q_1\delta q_2 \Big) \nn\\
&&-\fft{(n-2)\gamma\omega_{n-2}}{8\pi}\Big( q_1^2\delta \mu+\ft{3n-7}{2(n-2)}\,\mu\, \delta\big( q_1^2\big)  \Big)\,,
\eea
where the refining mass $\widetilde{M}$ is defined by
\be \widetilde{M}=M-\ft{\big(2(n-2)\gamma-(n-3) \big)\omega_{n-2}}{32(n-1)\pi}\fft{q_2^2}{\mu} \,.\ee
Then the Wald equation suggests the first law is
\bea\label{nonfirstlaw2}
d\widetilde{M}&=&T dS-\fft{(n-2)\gamma\omega_{n-2}}{8\pi}\Big(q_2 d q_1+\big(1-\ft{n-3}{2(n-2)\gamma} \big)q_1 d q_2 \Big) \nn\\
&&+\fft{(n-2)\gamma\omega_{n-2}}{8\pi}\Big( q_1^2 d \mu+\ft{3n-7}{2(n-2)}\,\mu\, d\big( q_1^2\big)  \Big)\,.
\eea
For our solution (\ref{nonsol2}), the temperature and Wald entropy are given by
\be T=\fft{n-3}{4\pi r_0}\,,\qquad S=\ft 14\mathcal{A} \Big( 1+\ft 12 \gamma q_1^2-\ft{\big(2(n-2)\gamma-(n-3) \big)q_2^2}{2(n-1)r_0^{2n-6}} \Big) \,.\label{nontementropy2}\ee
However, by plugging these results into the first law (\ref{nonfirstlaw2}), we find that it picks out a special coupling $\gamma=1$ when we impose the boundary condition that $A_{\bar a}$ vanishes on the horizon. In fact, relaxing the horizon condition for $A_{\bar r}$, we find in general $\delta H_+$ becomes non-integrable
\be\label{nonHh} \delta H_+=T\delta S+\fft{(n-3)(\gamma-1)\omega_{n-2}}{16(n-1)\pi}\Big(2(n-3)\,\mu\, \delta\big( q_1^2\big)+3(n-2)\,q_1^2\delta \mu \Big) \,.\ee
Thus, the Wald entropy formula is invalid for a generic coupling owing to existence of the non-integrable one form on the r.h.s of (\ref{nonHh}). These results are consistent with our discussions in section \ref{waldthermo}. Nonetheless, combining (\ref{nonHinfty}) and (\ref{nonHh}) and using the Wald equation one can formally write down a first law for the solution (\ref{nonsol2})
\bea\label{nonfirstlaw3}
d\widetilde{M}&=&T dS-\fft{(n-2)\gamma\omega_{n-2}}{8\pi}\Big(q_2 d q_1+\big(1-\ft{n-3}{2(n-2)\gamma} \big)q_1 d q_2 \Big)\nn\\
&&+\fft{\omega_{n-2}}{16(n-1)\pi}\Big[(n-2)\Big((5n-11)\gamma-3(n-3) \Big) q_1^2 d \mu \nn\\
&&\qquad \qquad\qquad+\Big( (5n^2-22n+25)\gamma-2(n-3)^2 \Big)\,\mu\, d\big( q_1^2\big)  \Big]\,.
\eea
This is a correct mathematic equation though its physical meaning is not so clear. Define a Legendre transformed energy function
\be E=\widetilde{M}+\fft{(n-2)\gamma\omega_{n-2}}{8\pi} q_1 q_2-\fft{(n-2)\omega_{n-2}}{16(n-1)\pi}\Big((5n-11)\gamma-3(n-3) \Big)\mu\, q_1^2  \,,\ee
the above first law simplified to
\be dE=T dS+\fft{(n-3)\omega_{n-2}}{16\pi}q_1 dq_2+\fft{(n-3)(n-\gamma)\omega_{n-2}}{16(n-1)\pi} \mu\, d\big( q_1^2\big)\,.\ee
In addition, we also find a Smarr-like relation
\be E=\fft{n-2}{n-3}T S+\fft{(n-3)\omega_{n-2}}{16\pi}q_1 q_2 \,,\ee
which is a natural result of the scaling dimensional arguments.

\subsubsection{Euclidean action}\label{newsection}

Since the non-integrability of $\delta H_+$ invalids the Wald entropy formula, we shall explore whether there exists an alternative approach to define the black hole entropy. It was first proposed in \cite{Gibbons:1976ue} that thermodynamic quantities for black holes can be calculated by means of {\it quantum statistical relation}:
\be F=I_{reg} T=\widehat{M}-T \widehat{S} \,,\label{free}\ee
where $F$ is the free energy, $I_{reg}$ is the regularized Euclidean action of black hole solutions and $\widehat{M}\,,\widehat S$ are black hole mass and entropy, respectively (they should not be confused with the mass and entropy defined from Wald formalism.).
The regularized Euclidean action can be defined by subtracting the action of a background solution with $\mu=0$ from the action of the black hole,
\be I_{reg}\equiv I_E[g_{\mu\nu}\,,A_\mu]-I_E[g^{(0)}_{\mu\nu}\,,A^{(0)}_\mu] \,.\ee
However, for our solution (\ref{nonsol2}) the limit $\mu\rightarrow 0$ is singular for a generic coupling $\gamma$. Instead, we derive a proper background solution by taking double scaling limit:
$\mu\rightarrow 0\,,q_2\rightarrow 0$ with $q_2^2/\mu\rightarrow \mathrm{const}$. The resulting expression for free energy is very simple
\be F= -\fft{(n-3)\omega_{n-2}\,q_2^2}{32\pi r_0} \,,\ee
In usual cases (such as a Schwarzschild black hole), one can derive both the mass $\widehat M$ and entropy $\widehat S$ independently as 
\be \widehat S=-\fft{\partial F}{\partial T}\,,\qquad \widehat{M}=F+T\widehat{S} \,,\ee
by making use of  Eq.(\ref{free}) and the first law $d \widehat M=T d \widehat S$.
However, for our solution, the first law (\ref{nonfirstlaw3}) is non-integrable. So we have to fix one of the two functions at first and derive the other one. We may take $\widehat M=M$ or $\widehat M=\widetilde{M}$. In both cases, we find the resulting entropy $\widehat S$ disagrees with the standard Wald entropy. Furthermore, if we instead require $\widehat S=S$, the mass $\widehat{M}$ will again disagree with $M$ and $\widetilde{M}$. As a matter of fact, the mass suffers from another shortcoming that it can not be connected to $M$ or $\widetilde{M}$ via a Lengendre transformation. This conflicts with the first law of thermodynamics. Hence, it is problematic whether $\widehat M$ has a correct thermodynamic meaning. These mismatches between Wald formalism and Euclidean method
imply that the thermodynamics of our solution deserves further investigations.

\section{ Non-minimal theory: case II }
Now let's consider the general non-minimally coupled theory described by
\be \mathcal{L}=R-\fft 14 F^2-\beta A^2 R+\gamma R_{\mu\nu} A^\mu A^\nu \,.\label{nonlagene}\ee
The maximally symmetric vacuum is Minkowski space-times. However, the Lorentz symmetry of the vacuum can break down because a constant vector is admitted as well, namely
\be ds^2=-dt^2+dr^2+r^2 d\Omega_{n-2}^2\,,\qquad A=q_1 dt \,.\label{nonflatback}\ee
Depending on the coupling constants, we find that there exist significantly different classes of asymptotically flat black hole solutions.

\subsection{Unconventional black hole}

The first class solution we find has an unconventional fall-off at asymptotic infinity. It reads
\bea\label{nonsol3}
&&ds^2=-f dt^2+\fft{dr^2}{f}+r^2 d\Omega^2_{n-2}\,,\nn\\
&&A=\sqrt{\ft{2(n-1)}{(n-3)(1-\gamma)}}\, f dt\,,\quad f=1-\fft{\mu}{r^{\fft{n-3}{2}}}\,.
\eea
provided the parametric relation
\be 2(n-1)\beta+(n-3)(\gamma-1)=0 \,.\ee
Here $\mu$ is an integration constant which should not be confused with the usual fall-off mode $1/r^{n-3}$.
This type solution was first found in \cite{Geng:2015kvs} for $\beta=\gamma/2$ and in \cite{Fan:2016jnz} for $\gamma=0$. Note that the reality of the vector requires $\gamma<1$ and the limit $\gamma\rightarrow 1$ or equivalently $\beta\rightarrow 0 $ is singular. In fact, the solution does not exist in the theory (\ref{simla2}) which has a single $\gamma$ term. As was shown in \cite{Fan:2016jnz}, the unusual fall-off $1/r^{\fft{n-3}{2}}$ in the metric functions corresponds to the longitudinal graviton mode, which is excited by the back-reaction effect of a background vector.

Since the solution has only one integration constant, we shall first analyze the general asymptotic solutions of the theory for generic coupling constants  before deriving the first law. Linearing the equations of motions around the background (\ref{nonflatback}), we find
\be\label{nonasymptotic2} A_t=q_1-\fft{q_2}{r^\sigma}+\cdots\,,\quad h=1-\fft{\mu}{r^\sigma}+\cdots\,,\quad f=1-\fft{\widetilde{\mu}}{r^\sigma}+\cdots \,,\ee
where $\sigma>0$ is an under-determined constant. The conventional solution has $\sigma=n-3$ and
\be \widetilde{\mu}=\fft{4\beta q_1 q_2+(1-\beta q_1^2)\mu}{1+\beta q_1^2} \,.\ee
 Thus, the general solution is characterized by three independent parameters $(\mu\,,q_1\,,q_2)$, as expected. However, the unconventional solution with $\sigma\neq n-3$ also exists provided
\be \widetilde{\mu}=\fft{\big( q_2+(2\beta-\gamma)\mu  q_1\big)\sigma}{2(n-2)\beta q_1} \,,\quad q_2=\fft{(2\beta-\gamma)(1+3\beta q_1^2)\mu q_1}{\big(4(2\beta-\gamma)+1\big)\beta q_1^2-1}\,,\ee
and the vector charge $q_1$ has been fixed as a function of $(n\,,\beta\,,\gamma)$ (the details is irrelevant in our discussion). Hence, one may worry about the existence of this type solutions since it needs a delicate fine tuning of the boundary conditions on the horizon\footnote{Perhaps, this type solutions is unstable because unlike conventional solutions, they in general do not have a convergent ADM mass. }. It is interesting that we do find such a solution when $\sigma=(n-3)/2$ (one can check that in this case $\widetilde{\mu}=\mu$ and the linearized solution (\ref{nonasymptotic2}) is just the exact solution (\ref{nonsol3}) ).

To understand the solution (\ref{nonsol3}) better, we develop its full large-$r$ expansions. We find\footnote{It was established in \cite{Geng:2015kvs} that for a special case $\gamma=\ft{n-3}{2(n-2)}$, the large-$r$ expansions are different from the equation (\ref{nonasymptotic3}). However, this does not change our conclusion. }
\bea\label{nonasymptotic3}
&&A_t=\sqrt{\ft{2(n-1)}{(n-3)(1-\gamma)}}\,\Big(1-\fft{\mu}{r^{(n-3)/2}} -\fft{m}{r^{n-3}}+\fft{\widetilde{a}_3}{r^{3(n-3)/2}}+\cdots\Big)\,,\nn\\
&&h=1-\fft{\mu}{r^{(n-3)/2}} -\fft{m}{r^{n-3}}+\fft{\widetilde{h}_3}{r^{3(n-3)/2}}+\cdots\,,\nn\\
&&f=1-\fft{\mu}{r^{(n-3)/2}} -\fft{2m}{r^{n-3}}+\fft{\widetilde{f}_3}{r^{3(n-3)/2}}+\cdots\,,
\eea
where $m$ is a new parameter associated with the condensate of the transverse gravitons. The higher order coefficients $\widetilde{a}_i\,,\widetilde{h}_i\,,\widetilde{f}_i$ can be solved in terms of functions of the two parameters $(\mu\,,m)$. Substituting (\ref{nonasymptotic3}) into the Wald formula, we obtain
\be \delta H_\infty=\fft{(n-2)\omega_{n-2}}{16\pi}\mu\, \delta \mu \,.\ee
Surprisingly, it does not receive any contributions from the transverse graviton mode! This is very different from the conventional black holes.
We define the mass as $\delta M_{uc}\equiv \delta H_\infty$, giving rise to
\be M_{uc}=\fft{(n-2)\omega_{n-2}}{32\pi}\mu^2 \,.\ee
For our solution (\ref{nonsol3}), the temperature and entropy are given by
\be T=\fft{n-3}{8\pi r_0}\,,\qquad S=\ft 14 \mathcal{A} \,.\ee
It follows that the first law and Smarr relation are given by
\be dM_{uc}=T dS\,,\qquad M_{uc}=\fft{n-2}{n-3}\,T S \,.\ee

To end this subsection, we point out that in the weak field limit the unconventional solution (\ref{nonsol3}) predicts a stronger gravitational force than the Schwarzschild black hole. For example, in the four dimension it has $1/r^{3/2}$-law rather than the well-known $1/r^2$-law. More interestingly, since the general asymptotic solution (\ref{nonasymptotic3}) has the usual fall-off mode $1/r^{n-3}$ as well, one can turn on or turn off the unconventional mode freely. This gives rise to new possibilities and candidates how the Newtonian inverse-squared law can be modified in galaxies and may be tested by observational data in astrophysics in the future.

\subsection{Stealth black hole and beyond}
Following the derivation in sec.\ref{sec4.2}, we find that there exists an exact stealth Schwarzschild black hole at the critical coupling $\gamma=\ft{n-3}{2(n-2)}$ whilst $\beta$ remains free. The solution reads
\bea\label{nonsol4}
&& ds^2=-fdt^2+\fft{dr^2}{f}+r^2 d\Omega^2_{n-2}\,,\quad  A_r=\sqrt{\ft{A_t^2}{f^2}-\ft{q_1^2}{f}}\,,\nn\\
&& A_t=q_1-\fft{q_2}{r^{n-3}}\,,\qquad f=1-\fft{2\mu}{r^{n-3}}\,.
\eea
We demand $q_1(\mu q_1-q_2)\geq 0$ to govern the reality of $A_r$. Formally, the solution is simply the one (\ref{nonsol2}) when $\gamma=\ft{n-3}{2(n-2)}$. Likewise, its first law can be studied along the discussions in sec.\ref{sec4.2}.

Surprisingly, if we fix the vector charge $q_1=1/\sqrt{\gamma-\beta}$, a new fall-off mode emerges in the metric function without altering anything else in the above solution. We find
\be f(r)=1-\fft{2\mu}{r^{n-3}}+\fft{\lambda}{r^\xi}\,,\quad \xi=\ft{2(n-2)\beta}{2\beta-\gamma}\,,\ee
where $\lambda$ is a new independent integration constant. The coupling constant $\gamma$ is still equal to $\ft{n-3}{2(n-2)}$ and $\beta<\gamma\,,\beta\neq \gamma/2$. For $\beta=0\,,\,\xi=0$ and for $\beta=-\ft 12(n-3)\gamma\,,\,\xi=n-3$. In both cases, the new mode is trivial and can be dropped in the metric function. When $\gamma/2<\beta<\gamma$ or $\beta<-\ft 12 (n-3)\gamma$, we find $\xi>(n-3)$, the $\lambda$ mode falls off faster than $1/r^{n-3}$ whilst for $-\ft 12 (n-3)\gamma<\beta<0$, we have $0<\xi<n-3$, the new mode falls off slower than the conventional one. Moreover, when $0<\beta<\gamma/2$, we have $\xi<0$, implying that the solution is no longer asymptotically flat, although the maximally symmetric vacuum of the theory is Minkowski space-times. In particular, when $\beta=\gamma/n$, we have $\xi=-2$, the solution becomes asymptotically (A)dS and the cosmological constant emerges as an integration constant, which is totally independent of the parameters in the Lagrangian density. It may be the first time to observe this phenomena in Einstein gravity except for conformal gravity.

Finally, it should be emphasized that the existence of such a new mode is peculiar for the theory we consider because the presence of both non-minimal couplings is essential to govern the existence of this type solution.

\section{Non-minimal theory: case III}

Now we study a certain non-minimally coupled theory which has $\beta=\gamma/2$ and also includes a bare cosmological constant and a bare mass term
\be \mathcal{L}=R-2\Lambda_0-\fft 14 F^2-\fft 12 m^2 A^2+\gamma G_{\mu\nu}A^\mu A^\nu \,.\label{lanongmass}\ee
The theory has been extensively studied in \cite{Geng:2015kvs,Chagoya:2016aar,Minamitsuji:2016ydr,Babichev:2017rti} but most of them are limited to the four dimension.

\subsection{Without Proca mass}
First, let's consider a simpler case $m^2=0$. From the discussions in above section, it is immediately to see that when the bare cosmological constant also vanishes, the asymptotically flat solutions (\ref{nonsol3}) and (\ref{nonsol4}) are still valid with the coupling constant $\gamma=\fft{(n-3)}{2(n-2)}$ in both cases.

For the same coupling constant, the solution (\ref{nonsol4}) can be generalized to non-asymptotically flat space-times when $\Lambda_0\neq 0$. We obtain\footnote{We do not find planar black holes in this case: $m^2=0\,,\Lambda_0\neq 0$. The reason is explained in next subsection.}
\bea && ds^2=-hdt^2+\fft{\sigma^2 dr^2}{h}+r^2d\Omega_{n-2}^2\,,\quad A_r=\fft{\sigma}{\sqrt{h}}\sqrt{\ft{A_t^2}{h}-q_1^2\sigma+\ft{4\Lambda_p(n-2)}{n-3}r^2}\,,\nn\\
     && A_t=q_0 r^2+q_1-\fft{q_2}{r^{n-3}}\,,\quad \sigma=\Lambda_p r^2+1\,,\quad h=g_4 r^4+g_2 r^2+1-\fft{2\mu}{r^{n-3}} \,,
\label{nonsol5}\eea
where various parameters are specified by
\be q_0=\fft{(n-3)\Lambda_p q_1}{n-1}\,,\qquad g_4=\fft{(n-3)\Lambda_p^2}{n+1}\,,\qquad g_2=\fft{2(n-3)\Lambda_p}{n-1}\,. \ee
Here $\Lambda_p$ is related to the bare cosmological constant
\be \Lambda_p=\ft{4\Lambda_0}{(n-3)\Big( (n-3)q_1^2-4(n-2) \Big)} \,. \ee
The solution in the $n=4$ dimension was first obtained in \cite{Chagoya:2016aar}. In the limit $\Lambda_0\rightarrow 0$, the solution reduces to (\ref{nonsol4}). Note that at asymptotic infinity, the solution (\ref{nonsol5}) does not approach neither
asymptotically (A)dS nor Minkowski space-times. We find
\be ds^2|_{r\rightarrow \infty} \rightarrow -\ft{(n-3)\Lambda_p^2}{n+1}r^4 dt^2+\ft{n+1}{n-3}dr^2+r^2 d\Omega^2_{n-2} \,,\ee
which is a $z=2$ Lifshitz space-times with conical singularities at infinity. As emphasized earlier, the reality of $A_r$ constrains the parameters space. We demand
\be \Lambda_0=0\,,\qquad q_1(\mu q_1-q_2)\geq 0 \,,\ee
or
\be \Lambda_0<0\,,\quad q_1^2\leq\ft{4(n-2)(n^2-1)}{(n-3)(n^2+1)}\,,\quad 2q_1 q_2\leq \mathrm{min}\Big\{ \mu q_1^2\,,\, \ft{(n-1)\mu}{n-3}\big( q_1^2-\ft{4(n-2)}{n-3} \big)   \Big\} \,.\ee

\subsection{With Proca mass}

With a nonzero Proca mass, it is of great difficult to solve exact black hole solutions in the theory (\ref{lanongmass}).  Interestingly, in \cite{Minamitsuji:2016ydr} the author found some exact solutions for certain coupling constants. Furthermore, in \cite{Babichev:2017rti} the authors developed a nice procedure to derive the general solution in the four dimension. Here we follow the discussions in \cite{Babichev:2017rti} and generalize the method to general dimensions.

\subsubsection{Derivation of the solutions}\label{nonderivation}
A neat observation in \cite{Babichev:2017rti} is that the equations $\mathcal{P}^r$ and $E_{rr}$ are purely algebraic for the metric function $f$ and the vector field $A_r$. Hence, they can be solved in terms of other functions and their derivatives immediately
\bea\label{algebraic}
&&f(r)=\fft{m^2r^2+(n-2)(n-3)\gamma k}{(n-2)\gamma\,\Big(r h'+(n-3)h \Big)}h(r)\,,\nn\\
&&A_r^2=-\ft{r^2}{2\Big( m^2r^2+(n-2)(n-3)\gamma k\Big)^2}\Big\{2(n-2)(m^2+2\gamma \Lambda_0)r\Big( \ft{h'}{h}+\ft{n-3}{r} \Big)  \nn\\
&&\qquad\,\, +\Big( m^2r^2+(n-2)(n-3)\gamma k \Big)\Big( \ft{A^{'2}_t}{h}+\ft{2(n-2)\gamma}{r}\big( \ft{A_t^2}{h} \big)' \Big)\Big\}\,.
\eea
The two equations (\ref{algebraic}) encode some universal information about the general solution. For example, requiring the metric functions behaves standard at asymptotically AdS space-times, namely at leading order $h=f=g^2r^2+\cdots$ at infinity, we find the effective cosmological constant should be proportional to the Proca mass squared
\be \Lambda_{\mathrm{eff}}=-\fft{m^2}{2\gamma}\,,\label{cosmconst}\ee
where the effective cosmological constant is parameterized by $\Lambda_{\mathrm{eff}}=-\ft 12 (n-1)(n-2)g^2$. On the other hand, the non-negativity of $A_r^2$ strongly constrains the parameters in the theory as well as those in the solution. For instance, at asymptotic infinity, $A_r^2$ behaves as (at leading order)
\be A_r^2=-\fft{(n-1)(n-2)(m^2+2\gamma\Lambda_0)}{m^4r^2}+O(1/r^4) \,,\ee
which implies that $\Lambda_0$ should not be bigger than $\Lambda_{\mathrm{eff}}$, namely
\be \Lambda_0\leqslant -\fft{m^2}{2\gamma}=\Lambda_{\mathrm{eff}} \,.\label{barelambda}\ee
These results are universal for the general solution.

To proceed our derivation, we parameterize the metric function $h$ and the vector field $A_t$ as
\be h=-\fft{2\mu}{r^{n-3}}+\fft{1}{r^{n-3}}\int \mathrm{d}r\,\Big(m^2 r^2+(n-2)(n-3)\gamma k\Big)H^{-2}(r)\,,\quad A_t=\fft{1}{r^{n-3}}\int \mathrm{d}r\, F(r)  \,.\label{nonGparametrize}\ee
It turns out that the remaining two independent equations $\mathcal{P}^t\,,E_{tt}$ are integrable for $H\,,F$ at the critical coupling constant $\gamma=\fft{(n-3)}{2(n-2)}$. First, the vector equation $\mathcal{P}^t$ simplifies to
\be \big( H F\big)'-\ft{n-4}{2r}H F=0\,, \ee
which can be solved immediately as
\be F=C_1 r^{\fft{n-4}{2}}H^{-1} \,,\label{nonGF}\ee
where $C_1$ is an integration constant. Its physically meaning will be explained later. Then the Einstein equation $E_{tt}$ reduces to a linear first order ODE of $H$ which is easy to integrate (we do not list it in the following due to its lengthy expressions). We obtain
\be H=\fft{C_2\Big( 2m^2 r^2+k(n-3)^2 \Big)r^{-\fft{n-4}{2}}}{2\Big((n-2)m^2-(n-3)\Lambda_0 \Big)r^2+2(n-2)(n-3)^2k-C_1^2} \,,\label{nonGH}\ee
where $C_2$ is a new integration constant. Substituting (\ref{nonGF}) and (\ref{nonGH}) into (\ref{nonGparametrize}), we can derive $h\,,A_t$ and then solve $f\,,A_r$ through the equation (\ref{algebraic}). This completes our derivation. The results depend on the topological parameter $k$ as well as the space-time dimension $n$. In the following, we will discuss the solutions in a case-by-case basis. Before this, we point out the above two integration constants $C_1\,,C_2$ are related to the vector charge $q_1$ and the effective cosmological constant, respectively. We find
\be C_1=\fft{(n-3)m\, q_1}{g\sqrt{n-1}} \,,\qquad  C_2=\fft{(n-2)m^2-(n-3)\Lambda_0}{m g\sqrt{n-1}} \,.\label{c12}\ee
Something else is we will not list the explicit expressions for $f$ and $A_r$ since they are lengthy and we have too many solutions in the following. The readers in need can easily find them through the equation (\ref{algebraic}).

\subsubsection{Planar black holes}
For simplicity, let us consider planar black hole solutions at first. For $n\neq 5$ dimensions, we obtain
\bea\label{massgenesol1}
A_t&=&q_1-\fft{q_2}{r^{n-3}}-\ft{(n-3)^3m^2 q_1^3}{2(n-1)(n-5)g^2\Big( (n-2)m^2-(n-3)\Lambda_0 \Big)r^2}\,,\nn\\
h&=&g^2r^2-\fft{(n-3)m^2q_1^2}{(n-2)m^2-(n-3)\Lambda_0}-\fft{2\mu}{r^{n-3}}\nn\\
&&+\ft{(n-3)^4m^4 q_1^4}{4(n-1)(n-5)g^2\Big( (n-2)m^2-(n-3)\Lambda_0 \Big)^2r^2}\,,
\eea
The event horizon is determined by the largest real root of $h(r_0)=0$ and its effective curvature is nonzero, given by
\be k_{\mathrm{eff}}=-\fft{(n-3)m^2q_1^2}{(n-2)m^2-(n-3)\Lambda_0}=-\fft{(n-3)q_1^2\Lambda_{\mathrm{eff}}}{(n-2)(\Lambda_{\mathrm{eff}}+\Lambda_0)} \,.\ee
The black hole mass $M$ is given by Eq.(\ref{mass}). Notice that the solution (\ref{massgenesol1}) is singular in the $n\rightarrow 5$ limit. As a matter of fact, a logarithmic term emerges in the original fall-off mode $1/r^2$. We find
\bea
&&A_t=q_1-\fft{q_2}{r^2}-\fft{m^2q_1^3\log{r}}{(3m^2-2\Lambda_0)g^2r^2} \,,\nn\\
&&h=g^2r^2-\fft{2m^2 q_1^2}{3m^2-2\Lambda_0}-\fft{2\mu}{r^2}+\fft{m^4q_1^2\log{r}}{(3m^2-2\Lambda_0)^2g^2r^2}\,.
\eea
 For above solutions, there is an another singular limit $\Lambda_0\rightarrow -\Lambda_{\mathrm{eff}}=\ft{(n-2)m^2}{n-3}$. However, in this case we do not find any physically interesting solutions. Notice that the above solutions do not have a regular limit for a vanishing Proca mass $m^2\rightarrow 0$. Thus, one cannot find planar black holes when $m^2=0$.

Evaluating the Wald formula for above solutions in general $n\geq 4$ dimensions, we obtain
\be \delta H_\infty=\delta \widetilde{M}-\fft{(n-3)\omega_{n-2}}{16\pi}q_2 d q_1 \,,\ee
where $\widetilde{M}$ is defined by
\be\label{improvedmass} \widetilde{M}=\ft{(n-2)\omega_{n-2}}{16\pi}\big(1+\ft{\Lambda_0}{\Lambda_{\mathrm{eff}}} \big)\mu \,.\ee
As discussed earlier, when $\beta=\ft 12\gamma$, $\delta H_+$ is given by Eq.(\ref{nonhorizon2}-\ref{improvedtementropy}). Then the Wald equation implies
the first law is
\be\label{firstnon3} d\widetilde{M}=\widetilde{T}  d \widetilde{S}-\fft{(n-3)\omega_{n-2}}{16\pi}q_2 d q_1 \,.\ee
For our solutions, the temperature $T$ and the quantity $\Phi$ are given by
\be T=-\ft{\Lambda_{\mathrm{eff}}\,r_0}{2(n-2)\pi}\Big( 1+\ft{(n-3)^2 q_1^2}{4(\Lambda_0+\Lambda_{\mathrm{eff}})r_0^2} \Big)\,,\quad \Phi=\gamma^{-1}\Big( -1+\ft{\Lambda_0}{\Lambda_{\mathrm{eff}}}+\ft{(n-3)^2 q_1^2}{4\Lambda_{\mathrm{eff}}\,r_0^2}\Big)   \,.\ee
It is easy to verify that the above first law is indeed satisfied. An open question is the physical interpretation of the improved temperature $\widetilde{T}$ in the thermodynamics. This is interesting and deserves further investigations in the near future. We also compute the regularized Euclidean action\footnote{For AdS solutions, one can also compute a renormalized Euclidean action using holographic renormalization. The result is slightly different from the above regularized Euclidean action, up to an additive constant. This is because there are some ambiguities in how to choose the background solution that should be subtracted from. Nonetheless, this little difference does not change our conclusions.} for our solution by subtracting the action of a background solution with $\mu=0$ and deduce the entropy (or mass). However, the results still suffer from the shortcomings that were found in sec.\ref{newsection}.


\subsubsection{Spherical black holes}
For spherically symmetric solutions, we obtain
\bea\label{massgenesol2}
A_t&=&-\fft{q_2}{r^{n-3}}-\fft{m^2q_1}{(n-3)(n-1)^2g^2\Big( (n-2)m^2-(n-3)\Lambda_0 \Big)}\times \\
&&\Big\{ (n-1)(n-3)\Big( m^2q_1^2-2(n-1)(n-2)g^2 \Big) -2r^2\times \nn\\
&&\Big( m^4q_1^2-(n-1)g^2\Big( (n-1)m^2+ (n-3) \Lambda_0  \Big)  \Big)\Big( 1-F\big(1\,,-\ft{n-1}{2}\,,-\ft{n-3}{2}\,,-\ft{(n-3)^2}{2m^2r^2} \big)  \Big) \Big\}\,,\nn\\
h&=&-\fft{2\mu}{r^{n-3}}+g^2r^2  F\big(1\,,-\ft{n-1}{2}\,,-\ft{n-3}{2}\,,-\ft{(n-3)^2}{2m^2r^2} \big)-\ft{(n-3)\Big(m^2 q_1^2-2(n-1)(n-2)g^2   \Big)}{(n-2)m^2-(n-3)\Lambda_0}\times\nn\\
&&\Big\{ F\big(1\,,-\ft{n-3}{2}\,,-\ft{n-5}{2}\,,-\ft{(n-3)^2}{2m^2r^2} \big)-\ft{(n-3)^3\Big(m^2 q_1^2-2(n-1)(n-2)g^2   \Big)}{4(n-1)(n-5)\Big((n-2)m^2-(n-3)\Lambda_0\Big)g^2r^2}  F\big(1\,,-\ft{n-5}{2}\,,-\ft{n-7}{2}\,,-\ft{(n-3)^2}{2m^2r^2} \big)  \Big\} \nn\,,
\eea
which looks quite complicated. At asymptotic infinity, the two functions behave as
\bea
&&A_t=q_1-\fft{q_2}{r^{n-3}}+\Big(\fft{\widetilde{a}_2}{r^2}+\fft{\widetilde{a}_4}{r^4}+\cdots\Big)\,,\nn\\
&&h=g^2r^2-\fft{2\mu}{r^{n-3}}+\Big(\widetilde{h}_0+\fft{\widetilde{h}_2}{r^2}+\fft{\widetilde{h}_4}{r^4}+\cdots\Big)\,,
\eea
where the dotted terms in brackets are infinite series of $1/r^2$ and all the coefficients $\widetilde{a}_i\,,\widetilde{h}_i$ are lengthy expressions of $n\,,\Lambda_0\,,m^2\,,q_1$. To be concrete, we give some lower lying examples
\bea
&&\widetilde{a}_2=\ft{(n-3)^3q_1\Big( (n-1)\big((n-2)m^2+(n-3)\Lambda_0 \big)g^2 -m^4 q_1^2 \Big)}{2(n-1)(n-5)\Big((n-2)m^2-(n-3)\Lambda_0  \Big)m^2g^2}\,,\nn\\
&&\widetilde{h}_0=\ft{(n-3)\Big( (n-1)\big(3(n-2)m^2+(n-3)\Lambda_0 \big)g^2 -2m^4 q_1^2 \Big)}{2\Big((n-2)m^2-(n-3)\Lambda_0  \Big)m^2}\,,\nn\\
&&\widetilde{h}_2=\ft{(n-3)^4\Big( (n-1)\big((n-2)m^2+(n-3)\Lambda_0 \big)g^2 -m^4 q_1^2 \Big)^2}{4(n-1)(n-5)\Big((n-2)m^2-(n-3)\Lambda_0 \Big)^2 m^4g^2}\,,
\eea
where $\widetilde{h}_0$ is the effective curvature of the horizon which in general is not equal to unity.

For even dimensions, the black hole mass is given by Eq.(\ref{mass})
whist for odd dimensions $n=2j+3\,,j=1\,,2\,,3\,,\cdots$
\be M=\fft{(n-2)\omega_{n-2}}{8\pi}\big(\mu-\ft 12 h_{2j} \big) \,,\label{massodd}\ee
which receives contributions from the vector field as well.

However, in the $n=5$ and $n=7$ dimensions, the solution $(\ref{massgenesol2})$ becomes singular. As a matter of fact, for the $n=5$ dimension we find
\bea
A_t&=&q_1-\fft{q_2}{r^2}+\ft{q_1\Big(4g^2\big(3m^2+2\Lambda_0\big)-m^4q_1^2 \Big)}{2m^2g^2\big( 3m^2-2\Lambda_0 \big)}\fft{\log{\big( m^2r^2+2 \big)}}{r^2} \,,\nn\\
h&=&g^2r^2+\ft{4g^2\big( 9m^2+2\Lambda_0 \big)-2m^4q_1^2}{m^2\big(3m^2-2\Lambda_0 \big)}-\fft{2\mu}{r^2}\nn\\
&&+\ft{\Big(4g^2\big(3m^2+2\Lambda_0\big)-m^4 q_1^2 \Big)^2}{2m^4g^2\big( 3m^2-2\Lambda_0  \big)^2}\fft{\log{\big( m^2r^2+2 \big)}}{r^2}\,.
\eea
The black hole mass is given by (\ref{massodd}) with
\be h_2=\ft{ \Big( 4g^2\big(3m^2+2\Lambda_0 \big)-m^4q_1^2 \Big)^2  }{2m^4g^2\big( 3m^2-2\Lambda_0 \big)^2}\log{\big( m^2 \big)} \,.\ee
In the $n=7$ dimension, we find
\bea
A_t&=& q_1+\ft{8q_1\Big(6g^2\big(5m^2+4\Lambda_0 \big)-m^4 q_1^2 \Big)}{3m^2g^2\big( 5m^2-4\Lambda_0 \big)r^2}-\fft{q_2}{r^4}\nn\\
&&-\ft{64q_1\Big(6g^2\big(5m^2+4\Lambda_0\big)-m^4q_1^2 \Big)}{3m^4g^2\big( 5m^2-4\Lambda_0 \big)}\fft{\log{\big( m^2r^2+8 \big)}}{r^4}\,,\nn\\
h&=&g^2r^2+\ft{12g^2\big( 15m^2+4\Lambda_0 \big)-4m^4q_1^2}{m^2\big(5m^2-4\Lambda_0 \big)}+\ft{16\Big(6g^2\big(5m^2+4\Lambda_0\big)-m^4 q_1^2 \Big)^2}{3m^4g^2\big( 5m^2-4\Lambda_0  \big)^2 r^2}\nn\\
&&-\fft{2\mu}{r^4}-\ft{128\Big(6g^2\big(5m^2+4\Lambda_0\big)-m^4 q_1^2 \Big)^2}{3m^6g^2\big( 5m^2-4\Lambda_0  \big)^2}\fft{\log{\big( m^2r^2+8 \big)}}{r^4}\,.
\eea
The black hole mass is still given by (\ref{massodd}) with
\be h_4=-\ft{ 256\Big( 6g^2\big(5m^2+4\Lambda_0 \big)-m^4q_1^2 \Big)^2  }{6m^6g^2\big( 5m^2-4\Lambda_0 \big)^2} \log{\big( m^2 \big)}\,.\ee

Notice that for all these solutions the limit $m^2\rightarrow 0$ is singular. Thus, one cannot recover the solution (\ref{nonsol5}) by simply sending $m^2\rightarrow 0$ from these solutions. Instead, one should follow the derivation in sec.\ref{nonderivation} (there the integration constants $C_{1\,,2}$ in (\ref{c12}) become singular in the limit $m^2\rightarrow 0$ and should be chosen properly again) and then re-derive the solution (\ref{nonsol5}).

Finally, for all the solutions above, the first law is still given by Eq.(\ref{firstnon3}) with $\widetilde{M}$ defined by Eq.(\ref{improvedmass}), even for odd dimensional solutions whilst the temperature $T$ and $\Phi$ are given by
\be T=-\ft{\Lambda_{\mathrm{eff}}\,r_0}{2(n-2)\pi}\Big( 1+\ft{(n-3)^2 q_1^2-4(n-2)(n-3)}{4(\Lambda_0+\Lambda_{\mathrm{eff}})r_0^2} \Big)\,,\quad \Phi=\ft{(n-3)^2q_1^2-4(\Lambda_{\mathrm{eff}}-\Lambda_0)r_0^2}{2\gamma\big(2\Lambda_{\mathrm{eff}}\,r_0^2-(n-2)(n-3) \big)}   \,.\ee
This is expected since the Wald formula given in Eq.(\ref{waldformula1}-\ref{waldformula3}) do not explicitly depend on the topology of the space-times.

\section{Conclusions}
In this paper, we study generalized Einstein-Proca theories in general dimensions by introducing either a quartic self-interaction term for the vector or non-minimally coupled terms between the curvature and the vector. In general, the gauge symmetry of the vector is explicitly breaking but can be restored at the linear level around any Ricci-flat metric, depending on the parameters of the theories.

We find that there are two distinct class solutions, both of which are general static and have maximal symmetry, depending on whether $A_r$ vanishes. In particular, the solutions with a nonzero $A_r$ have some attractive features that we do not find for the solutions with a vanishing $A_r$. The first is in many cases (for example the minimal theory which is simply the Einstein-Proca theory extended with a quartic self-interaction term for the vector field) we can analytically solve all the equations of motions and exactly obtain the general static maximally symmetric black hole solutions. This is quite surprising since it is known that the Einstein equations are highly non-linear and one has not found any analytical solutions in the standard Einstein-Proca theory. The underlying reason is the equations $\mathcal{P}^t\,,E_{tt}$ are purely algebraic for the metric function $f$ and the vector component $A_r$, which can be immediately solved in terms of functions of $h\,,A_t$ and their derivatives. It turns out that under certain parametrizations of $h\,,A_t$, the remaining independent equations $\mathcal{P}^t\,,E_{tt}$ are greatly simplified to first order ODEs, which are easy to integrate.

Second, the reality of $A_r$ provides strong constraints on the parameters of the solutions as well as those in the Lagrangian density.
Third, we adopt the Wald formalism to derive the first law of thermodynamics for all of the solutions. However, the situation is subtle for the solutions with a nonzero $A_r$ because to govern the validity of Wald entropy formula, we need impose proper boundary conditions that the local diffeomorphism invariant of the vector $A_{\bar a}$ vanishes on the horizon, which unfortunately turns out to be too strong for this type solutions. The reason is $A_r$ does not have corresponding vector charges since it is a purely algebraic degree of freedom. Thus, we have to relax the horizon condition for $A_{\bar r}$ but this conversely results to a non-integrable $\delta H_+$ which invalids the Wald entropy formula. The thermodynamics of such solutions deserves further studies.

Finally, we also obtain some exact black hole solutions with vanishing $A_r$. In particular, one of the solutions has an unconventional fall-off mode, which is interpreted as the longitudinal gravitons excited by the vector field. In the weak field limit, the solution has a stronger gravitational force than the usual Newton's $1/r^2$-law. This is particularly interesting in astrophysics since it provides new candidates to modify the Newton's inverse-squared law.

\section*{Acknowledgments}
I am grateful to Xing-Hui Feng and Hong L\"u for their participation in the early stage of this work. Z.Y. Fan is supported in part by NSFC Grants No. 11273009 and No. 11303006 and also supported by Guangdong Innovation Team for Astrophysics(2014KCXTD014).


\end{document}